\documentclass{article}
\usepackage{graphicx}
\begin{document}

\Large {\bf Is CeNiSn a Kondo semiconductor? - break-junction
experiments}

\vspace{1cm}
\large
Yu.~G.~Naidyuk$^{1}$\footnote{naidyuk@ilt.kharkov.ua}, K.
Gloos$^2$, and T. Takabatake$^3$

\vspace{1cm}
 {\it $^1$ B.~Verkin Institute for Low Temperature Physics and
  Engineering, NAS of Ukraine, 61164 Kharkiv, Ukraine

$^2$ Institut f\"ur Festk\"orperphysik,  Technische Universit\"at
Darmstadt, D-64289, Germany and  Department of Physics,
University of Jyv\"askyl\"a, FIN-40351 Jyv\"askyl\"a, Finland

$^3$ Department of Quantum Matter, ADSM, Hiroshima University,
Higashi-Hiroshima 739-8526, Japan}

\vspace{1cm}
\begin{abstract}
We investigated break junctions  of the Kondo semiconductor
CeNiSn, both in the metallic and in the tunnelling regime, at low
temperatures and in magnetic fields up to $8\,$T. Our experiments
demonstrate that direct CeNiSn junctions have typical  metallic
properties  instead of the expected semiconducting ones. There is
no clear-cut evidence for an energy (pseudo)gap. The main spectral
feature, a pronounced $\sim 10-20\,$ meV wide zero-bias
conductance minimum, appears to be of magnetic nature.
\end{abstract}

 \normalsize \vspace{1cm}
 Cerium intermetallic compounds can have different
ground states, depending on the hybridization between f- and
conduction electrons. CeNiSn is usually classified as a Kondo
semiconductor, originally because of the enhanced electrical
resistivity at low temperatures  \cite{Takab90}. But when high
quality samples became available, the low-temperature resistivity
turned out to be metallic \cite{Nakamo95}. Tunnel spectroscopy
provides a direct access to the electronic density of states
(EDOS) \cite{Wolf85}. Using mechanically controllable break
junctions (MCBJ), Ekino et al. \cite{Ekino95} observed d$I/$d$V$
spectra with $\sim 10\,$meV broad zero-bias (ZB) minima. They
assumed -- without further experimental evidence -- that their
junctions were in the tunnel regime, and interpreted the ZB
minima as being due to a gap in the EDOS. These ZB minima were
found to be suppressed in magnetic fields $B\geq$14\, T only
along the a-axis, indicating as a crossover from a pseudogap to a
metallic heavy-fermion state \cite{Davydov97}. Our investigation
of MCBJs of CeNiSn, both in the metallic (direct contact) and in
the vacuum-tunneling regime, is based on three CeNiSn single
crystals with long sides in the a, b, and c-direction of the
orthorhombic crystal lattice, respectively. Magnetic fields up to
8 T could be applied perpendicular to the long side of the sample
(perpendicular to current flow). For further details see
Refs.~\cite{Gloos99,Naid00}.
  \begin{figure}[t]
 \begin{center}
   \includegraphics[width=10.5cm]{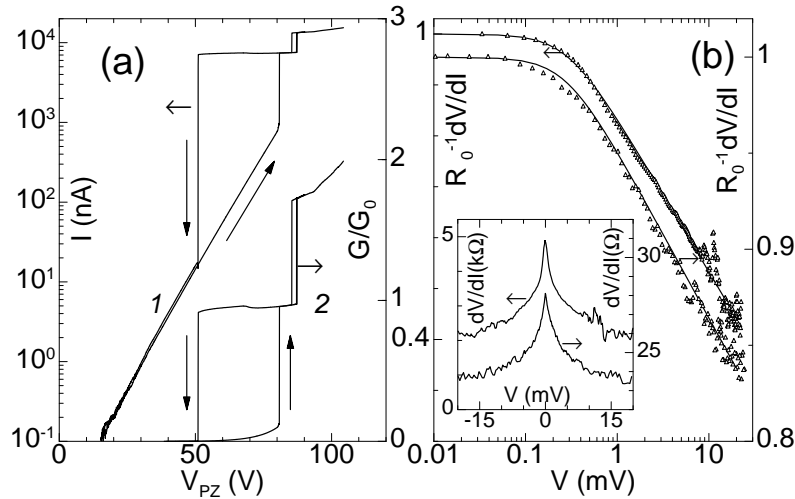}
  \caption[]{(a) Current $I$ through a CeNiSn MCBJ in c-direction
    vs piezo voltage $V_{\rm{PZ}}$ (curve 1) at $T=0.1\,$K and
    $V_{\rm{bias}}$=0.1\,V.  Curve 2   shows the same data in
   conductance units, normalized to  the quantum conductance
   G$_0 = $2e$^2/h \simeq 77.5\,\mu$S.
  (b) Reduced d$V$/d$I$ vs $V$ for  two contacts  with
  low (28\,$\Omega$) and high (5\,k$\Omega$) resistance. Solid lines
  are fits to the Daybell formula
  \cite{Daybell73} $R=1-A\,\log(1+(V/V_0)^2)$, with A= 0.019(0.066),
  $V_0=0.357(0.345)\,$mV for the bottom(top) curve. Inset shows the same
  curves for both polarities.}
  \label{cns1}
  \end{center}
  \end{figure}

To identify the regime of charge transport through the junctions
we measured how the contact resistance depends on the distance
between the two broken pieces of the sample, set by the piezo
voltage $V_{\rm PZ}$. Fig.~\ref{cns1}a clearly shows an
exponential $I(V_{\rm PZ})$ dependence at constant bias voltage
 as long as $R>100\,$k$\Omega$, as expected for true
vacuum tunneling. The step-like change of conductance at $R<
$G$_0^{-1}=$h$/$2e$^2 \simeq 13\,$k$\Omega$ characterizes
atomic-size metallic contacts \cite{Krans95}. Therefore, our
CeNiSn junctions with $R<13\,$k$\Omega$ are made up of  a metallic
constriction.

\begin{figure}[t]  \begin{center}
  \includegraphics[width=10.5cm]{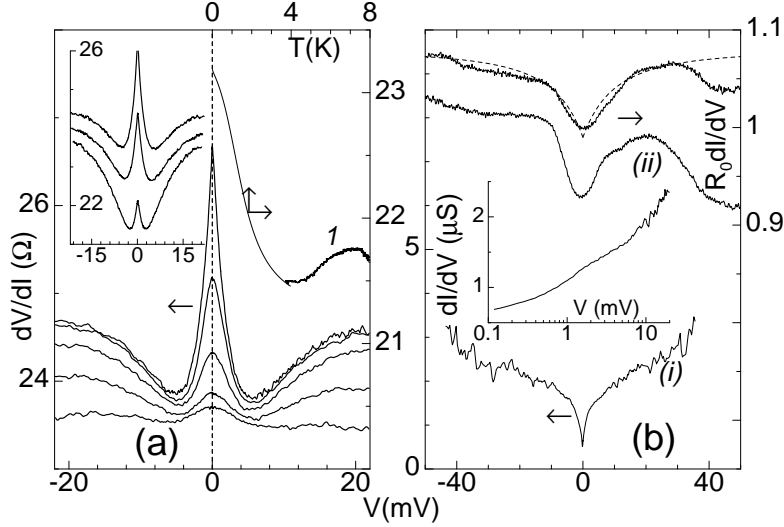}
  \caption[]{(a) d$V$/d$I$ of a metallic CeNiSn MCBJ in b-direction at $T$(K)=0.1,
    2, 4, 6 and 8 (top to bottom), and the temperature dependence
   of the ZB  resistance (curve 1). (The ZB resistance is  smaller
    because the junction changed slightly after heating to 8\,K).
    Inset:  d$V$/d$I$ vs $V$ of this  contact at $T = 0.1\,$K and
    $B$(T)=0, 4, and 8 (top to bottom).
     (b)  Two type of (see text) d$I$/d$V$ vs $V$ of CeNiSn tunnel junctions
     in a- (i) and c-direction (ii) at $T=0.1\,$K, $B=0$.
     The upper d$I$/d$V$  is (ii) curve ($R_0\simeq$1\,M$\Omega$)
     at $B=6\,$T. The dashed  line describes qualitatively the
     results of Coulomb blockade, according to Eq.(36) from \cite{Gloos99}.
     The inset shows  the bottom  curve  on a log-plot.}
  \label{cns2}
 \end{center}
  \end{figure}

Metallic MCBJs show rather similar spectra with a pronounced ZB
peak (Fig.~\ref{cns1}b, inset). d$V$/d$I$ decreases
logarithmically between 1 and 10 mV. It can be well described by
an empirical formula for the temperature dependence of Kondo
scattering if $T$ is replaced by $V$ (Fig.~\ref{cns1}b  ).
Fig.~\ref{cns2}a shows the temperature and field dependence of the
d$V$/d$I$ spectra in b-direction. In addition to the ZB peak, the
background of the curves increases with voltage, resulting in a
double-minimum structure similar to that found earlier
\cite{Ekino95,Davydov97}.
 However, contacts with $R < 13\,$k$\Omega$ are not in the tunnel regime,
and those ZB anomalies can not be attributed directly to the gap
in the EDOS, as proposed in Refs.~\cite{Ekino95,Davydov97}.
According to Fig.~\ref{cns2}a, the d$V/$d$I(V)$ spectra look like
the $T$-dependence of the ZB contact resistance d$V/$d$I(T,V=0)$.
This characterizes the thermal and not the ballistic regime of
metallic contacts \cite{Naid98}. A natural explanation for the ZB
peak in Fig.~\ref{cns2}a could be Kondo scattering which is indeed
supported by fitting of d$V$/d$I(V)$ with the corresponding
expression, see Fig.~\ref{cns1}b. The enhanced  low-$T$ resistance
probably has the same origin as that observed earlier on less
perfect (less pure) CeNiSn samples \cite{Takab90}. Probably the
quality of the interface is degraded with respect to the bulk
material, for example due to  mechanical stress.

Two different type of CeNiSn tunnel junctions can be distinguished
(Fig.~\ref{cns2}b): $(i)$ Contacts with a large ($>100\%$) ZB
minima, similar to the d$V$/d$I$ - maxima of the metallic contacts
in Fig.~\ref{cns1}b. $(ii)$ Contacts with a shallow ($\sim10\%$)
minimum. The latter have a relatively  broad and also more
asymmetric ZB dip. A magnetic field only slightly broadens ZB
minima. If we attribute those ZB minima to a gap of the EDOS, then
its width determined by the position of the maxima is $2\Delta
\sim 20\,$mV. At a characteristic temperature of $T_{\rm{c}}
\approx 10\,$K this yields an excessibly large $2\Delta/$k$_{\rm
B}T_{\rm c} \sim 20$. But there are other explanations, too. The
first type of spectra could be caused by magnetic scattering. For
example, evaporating less than one monolayer of magnetic
impurities onto thin film metal-oxide-metal planar tunnel
junctions can produce either a ZB conductance maximum or a
minimum, depending on the sign of the exchange integral between
conduction electron spin and magnetic impurity spin \cite{Wolf85}.
The size of those anomalies is of order 10\%. A giant ZB
resistance maximum similar to that in Fig.~\ref{cns2}b and with a
logarithmic variation between a few mV and 100 mV was observed in
Cr-oxide-Ag tunnel junctions, and explained by Kondo scattering as
well \cite{Wolf85}.

Another explanation could be Coulomb blockade, depending on the
capacitance of the tunnel junctions. Pronounced ZB minima could
result when the junctions consist of several isolated metallic
(magnetic) clusters, formed accidentally while breaking the
sample. Their capacitances are not shortcircuited by the
distributed lead capacitances, therefore Coulomb blockade can be
much stronger than at solitary junctions.

In summary, MCBJ experiments so far do not provide clear-cut
evidence for an energy (pseudo)gap of CeNiSb, even when the
junctions are in the true vacuum tunnel regime. The observed
anomalies could equally well be produced by Kondo scattering or
even by Coulomb blockade.

\end{document}